\documentstyle[preprint,eqsecnum,prb,aps]{revtex}
\newcommand{\bq}{\begin{equation}}
\newcommand{\enq}{\end{equation}}
\newcommand{\bqa}{\begin{eqnarray}}
\newcommand{\enqa}{\end{eqnarray}}

\begin{document}
\title{Semiclassical theory of  excitonic polaritons in  a planar semiconductor microcavity  }
\author{ Gu Xu,Dingzhou Li\cite{byline},Bingshen Wang$^\dag$, Zhao-bin Su}
\address{Institute of Theoretical Physics, Academia Sinica, Beijing, 100081, China}
\address{$^\dag$  Institute of Semiconductors, Academia Sinica, Beijing, 100083, China}
\date{\today}
\maketitle
\begin{abstract}
We present a comprehensive theoretical description of quantum well exciton-polaritons imbedded in  a planar semiconductor microcavity. The exact non-local  dielectric response of the quantum well exciton is  treated  in detail. The 4-spinor structure of the hole subband in the quantum well is considered,   including the pronounced band mixing effect. The scheme is self-contained and can be used to treat different  semiclassical aspects of the microcavity properties. As an example, we analyze the "selection" rules for the exciton-cavity mode coupling for different excitons.   
\end{abstract}
\preprint{}
\abovedisplayskip=3pt plus 1pt minus 1pt
\belowdisplayskip=3pt plus 1pt minus 1pt
\parskip=0pt plus.5pt minus 0.5pt
\section{Introduction}
Since the first experiment on  quantum well (QW) excitonic polaritons in the semiconductor microcavity (SMC)\cite{weisbuch92}  extensive experimental and theoretical studies\cite{weisbuch941} \cite{Stanley}  have been conducted  on excitonic polaritons in  QW's and multiple QW's embedded in an SMC, and even  bulk excitonic polaritons in an SMC. The SMC exciton-polariton was first described by so called linear dielectric model\cite{zhu} which was originally proposed for atoms in  a microcavity. Since then, most analysis on the microcavity problem has been  more or less based upon the assumed analogy between  excitons and atoms. The non-locality of the dielectric response of the exciton in a QW and the intricacy of real excitons due to the complex structure of the valence band has not been treated seriously.
\par
In fact, excitons differ from an atomic excitation in that they are a sort of excitation from electrons in the valence band confined by the QW's barriers. The excitonic polariton is essentially a quantum many-body effect and can be understood to be a result of the medium's polarization. In the calculation of  the dielectric response to light propagation in a QW, we have translational invariance only in the plane perpendicular to the growth direction,  so the nonlocality  of the dielectric response along the growth direction should be carefully treated. This makes the problem of a QW exciton embedded in an SMC non-trivial\cite{tasson90}. In particular, the barrier confinement effect in association with the 4-component spinor-like hole wave function should appear in the nonlocal dielectric response. The confinement induces hybridization between heavy and light hole subbands, which sometimes \cite{sham}, \cite{ycchang} plays a nontrivial role at the top region ($\Gamma$-point) of the valence band and has interesting observable consequences for the optical transitions of the excitons in a QW\cite{zhubf1}-\cite{andreani2}. This makes the non-local dielectric response of the QW more complex. To our knowledge, the non-locality of  QW exciton-polariron has been considered\cite{tasson90} for  the simplified exciton model, while the real QW exciton model including non-locality and subband mixing effects has not yet been considered  seriously, especially in a microcavity environment. On the other hand, although the effect of such hybridization induced exciton is not very strong,  the exciton-photon coupling is greatly enhanced in the microcavity. Thus it is interesting to  investigate the nature of such hybridization induced  QW excitonic polaritons in the SMC environment.\par
Based upon the above considerations, in this paper we provide a self-consistent semiclassical description of the excitonic
polaritons in a QW embedded in  an SMC, in which the hole subband hybridization, the nonlocality induced by the QW barrier confinement and the
boundary conditions for the SMC in connection with distributed Bragg reflectors(DBR's) are all
 consistently taken into account. The light field is treated semiclassically
but the  polarization of the medium in the QW is treated in the context of
quantum many-body theory with the electromagnetic wave-electron-hole
interaction, so that the description of the effective coupling between light and excitons does not need any phenomenological input ``coupling constant''. The only input parameters are the width of the exciton levels. The calculated Rabi splitting of  conventional excitonic polaritons could be considered as an improvement to the so-called dielectric model in which the photon-exciton
coupling constant is estimated  using the electric field of a cavity mode and  and the oscillator strength of an exciton. We will show also the
complexity in introducing such an effective oscillator strength due to
the coupling of the motion along or perpendicular to the growth
directions for the hybridization induced polaritons. As an interesting
effect of the barrier confinement induced electron and hole envelope
wave functions for a system with a pair of symmetric DBR's, we apply 
parity symmetry analysis along the growth direction to the  DBR's 
and the SMC confined QW, and obtain a kind of selection rule for  excitonic
polaritons in the well, which is also quite useful for  finding the hole
subband hybridization induced polaritons. As an  application of our scheme, we have calculated the reflection spectrum for several different excitons and for different incident angles,  and have obtained some interesting results, which  will be published elsewhere.  Here we only discuss the  formal aspects of our approach. 
\par
Since this topic is in an interdisciplinary field between  condensed
matter physics and quantum optics, we  present our discussion in a self-contained way for readers in both fields. This paper
is organized as follows. In Sec.~II., we formulate the description of the excitonic polaritons in a QW embedded in an SMC in terms of a propagating
electromagnetic field, while the SMC is further confined by a pair of DBR's.
The basic spirit of its classical electrodynamic aspect mainly follows ref. \cite{liu},  but the crucial difference is in the physics
contained in the formal expression for the dielectric response. Reference \cite{liu} is valid chiefly for the intra-band transition while  our paper  is devoted to the interband transition,  particularly  the hybridization effect of the hole band. Sec.~III is devoted to a quantum many-body description for the medium
polarization in a QW, in which the hybridization for the heavy hole subband and
the light hole subband is emphasized. Although the complex valence band in a QW and the related exciton problem has been  widely treated in the literatures,  a complete treatment that starts from the quantum many-body theory for the QW exciton,  and includes non-local response function and complete symmetry analysis for the 4-component spinor like wavefunction, is still absent. We shall present  such a complete and detailed treatment in this section. In Sec.~III-A, we give an expression for the medium polarization propagator with both positive and negative
frequency parts, which is applicable to generic non-translational
invariant systems. In Sec.~III-B, by summing over the Kramer's degenerate
states, we show the detailed expression for the nonlocal conductivity tensor
which is proved to be diagonal and can be factorized into a sum of bi-products.
This makes its non-locality effect  explicit. In Sec.IV we present the final complete set  of self-consistent equations as well as the boundary conditions which are put into a discrete version for the convenience of practical calculation. As an application, we discuss the "selection rules" for the  exciton-cavity  coupling.  Sec.V gives some concluding remarks.

\section{Description of a Quantum Well in a Cavity Confined
by Distributed Bragg Reflectors.}
The SMC under consideration consists of a
QW of thickness $\Lambda$ embedded in a thin layer
semiconductor material which is sandwiched between a pair of
distributed Bragg reflectors. This semiconductor layer
serves as the barrier for the electrons and holes in the QW, while it forms an optical cavity confined by the DBR's. The whole structure is schematically illustrated  in Fig.~\ref{MC}. Although we only treat the case for one QW in the cavity, generalization to more complex cases can be easily done.  The thickness of the
cavity is $L_c$ with $\varepsilon_c(\omega)$ as its medium's dielectric
constant, where $\omega$ is the light radiation frequency. We assume that the QW medium has the same (background) dielectric constants  as those  of the cavity
medium. The left (right) DBR is constructed from $N_L$ $(N_R)$
pairs of two alternating layers. One of the two layers has a length  $L_1$ and dielectric constant  $\varepsilon_1(\omega)$ while
the other  $L_2$ and $\varepsilon_2(\omega)$, respectively.
We choose the z-axis as the growth direction with $z=0$ being the 
center of the system. Thus,  the QW is located at
$(-\frac{\Lambda}{2},\frac{\Lambda}{2})$ and the SMC occupies a
region from $-\frac{L_c}{2}$ to $\frac{L_c}{2}$ with
$L_c\ge\Lambda$. Consequently, $-\frac{L_c}{2}$
$(\frac{L_c}{2})$ is the boundary between the left (right)
DBR and the SMC. Moreover, we rotate the coordinate axis around
the $z$-axis in such a way to make the incident light propagate in
the $x-z$ plane. Then the wave vector has the form 
$\vec{q}=(q_x\equiv q_\parallel$, 0, $q_z\equiv q_\perp)$ with $q_y$
being always equal to zero. As a result, for the $p$-polarized wave,
the electric field has only $x-$ and $z-$ components, $E_x$ and $E_z$, respectively, 
while for the $s$-polarized wave the electric field has only a
$y$-component $E_y$. For a propagating wave with fixed frequency and
$\vec{q}_\parallel=q_\parallel\vec{e}_x$ (where
$\vec{e}_\mu,~\mu=x,y,z$ is the unit vector along the $\mu$-axis),
its electric field has the form 
\bq
\vec{E}(\vec{r},t)=\vec{E}(\omega,\vec{q}_\parallel,z)e^{i(q_\parallel
x-\omega t)}=\vec{E}(\omega,\vec{r})e^{-i\omega t}.
\label{m1}
\enq
The Maxwell equation for the electric field can be written as
\bq
\Bigl[\tensor{I}\;\Bigl(\nabla^2+
\frac{\omega^2}{c^2}\varepsilon_i(\omega)\Bigr)-\vec{\nabla}
\vec{\nabla}\Bigr]\cdot\vec{E}(\omega,\vec{r})=-\frac{4\pi
i \omega}{c^2}\vec{j}(\omega,\vec{r}) \label {m2} 
\enq
in which ``i'' is the medium index. Taking into consideration  Eg.(\ref{m1}), Eq.(\ref{m2}) can be simplified to
\bq
\tensor{L}_i(\omega,\vec{q}_\parallel)\cdot\vec{E}(\omega,\vec{q}_\parallel,z)=-\frac{4\pi i \omega}{c^2}\vec{j}(\omega,\vec{q}_\parallel z) 
\label{m3}
\enq
with
\bq
\tensor{L}_i(\omega,\vec{q}_\parallel)=\tensor{I}\Bigl(\frac{\partial^2}{\partial z^2}-q^2_\parallel+\frac{\omega^2}{c^2}q_i(\omega)\Bigr)-\Bigl(i\vec{q}_\parallel
+\vec{e}_z\frac{\partial}{\partial z}\Bigr)\Bigl(i\vec{q}_\parallel
+\vec{e}_z\frac{\partial}{\partial z}\Bigr). \label{m4}
\enq
For simplicity, from now on we discuss mainly the
$p$-polarized wave propagation. It is straightforward to convert the
discussions for a $p$-polarized wave into those for an $s$-polarized wave.
We will do this when necessary.
Outside the QW medium, the polarization current density
$\vec{j}(\omega,\vec{q}_\parallel,z)=0$ in Eq.~(\ref{m3}). We then have plane wave
solutions with 
\bq
q_{\perp,i}=\Bigl[\frac{\omega^2}{c^2}\varepsilon_i(\omega)-q^2_\parallel\Bigr]^{1/2}
\label{m5}\enq
For the left DBR, denoting the $x$-components of the incident and
reflected amplitudes of the electric field as $A_0$ and $B_0$, which
are defined as  approaching to and leaving from the right boundary of the left DBR. By applying the method of ref.~\cite{yariv}, we obtain the corresponding electric field amplitude at the right boundary of
the left DBR as $A_1$ and $B_1$,  which are also defined by approaching the boundary from the left as
\bq
\pmatrix {A_0\cr B_0} =\pmatrix{ T^L_{11}\,, & T^L_{12}\cr 
T^L_{21}\,, & T^L_{22}} \pmatrix{ A_1\cr  B_1}
\label{m6} \enq
with
\bq
\pmatrix {T^L_{11}\,, & T^L_{12}\cr  T^L_{21}\,, & T^L_{22}} =
\pmatrix{ t^{ L}_{11}\,, & t^{ L}_{12}\cr  t^{ L}_{21}\,, 
& t^{ L}_{22}} ^{N_L},
\enq
where ${\bf T}^L$ is the transfer matrix of the left BDR, ${\bf t}^L$ is the tranfer matrix of a pair of layers with different dielectric constants, as a unit cell of the left DBR, in the DBR and $N_L$ is the number of the pairs in the DBR. Then for the right DBR, we introduce further $A_3$ and $B_3$, the forward (along the $z$-direction) and backward (in the negative $z$ direction) amplitudes of the electric field, respectively. They are defined by approaching from the right to the left boundary of the right DBR. Meanwhile, $A_4$ and $B_4$ are the corresponding amplitudes
at the right boundary of the right DBR which are also defined by
approaching the boundary from the right. We can similarly define the
transfer matrix by solving the homogeneous Maxwell equations \cite{yariv} to  obtain
\bq
\pmatrix {A_3 \cr B_3}=\pmatrix {T^R_{11}\,, & T^R_{12}\cr 
T^R_{21}\,, & T^R_{22}} \pmatrix {A_4 \cr  B_4 } \label{m8}\enq
with
\bq
\pmatrix {T^L_{11}\,, & T^R_{12}\cr  T^R_{21}\,, & T^R_{22}} =
\pmatrix {t^{ R}_{11}\,, & t^{ R}_{12}\cr  t^{ R}_{21}\,, 
& t^{ R}_{22}} ^{N_R}
\label{m9}\enq
The detailed expressions for the matrix elements $t^{ L}_{i,j}$
and $t^{ R}_{ij}$ with $i,j=1,2$ are shown in Appendix A.
\par

\par
In the QW region, there is a sort of additional  polarization
effect of the medium other than that described by its background dielectric
constant. This is because the virtual electron-hole pairs
creation and annihilation processes will renormalize the light
propagation in the QW region. In particular, the Coulomb interaction
between the virtual electron-hole pairs may play a crucial role in a 
certain frequency range. In this section, we just introduce formally
a conductivity tensor ${\sigma}(\omega,\vec{q}_\parallel;z,z')$ to
describe this effect, such that
\bq
\vec{j}(\omega,\vec{q}_\parallel,z)=
\int\!d z'\tensor{\sigma}(\omega,\vec{q}_\parallel;
z,z')\cdot\vec{E}(\omega,\vec{q}_\parallel,z')
\label{m10}\enq
which will be further investigated in the next section. Therefore, in
the SMC region (including the QW), the Maxwell equation (\ref{m3}) and
(\ref{m4}) can be converted into an integral equation as
\bqa
\vec{E}(\omega,\vec{q}_\parallel;z)=& \vec{E}^{(c)}(\omega,\vec{q}_\parallel;z)
-\frac{4\pi i\omega}{c^2}
\int^{\Lambda/2}_{-\Lambda/2}\!d z'\int^{\Lambda/2}_{-\Lambda/2}d z''
\tensor{G}(\omega,\vec{q}_\parallel;z,z')\cdot \cr 
& \tensor{\sigma}(\omega,\vec{q}_\parallel;
z',z'')\cdot\vec{E}(\omega,\vec{q}_\parallel,z'')
\label{m11}\enqa
in which
\bq
\vec{E}^{(c)}(\omega,\vec{q}_\parallel;z)=(A_c e^{i q_{\perp,c}z}+
B_c e^{-i q_{\perp,c}z})\vec{e}_x-\frac{q_\parallel}{q_{\perp,c}}
(A_c e^{i q_{\perp,c}z}-B_c e^{-i q_{\perp,c}z})\vec{e}_z
\label{m12}\enq
is the homogeneous solution of Eq.(\ref{m3}), while
${G}(\omega,\vec{q}_\parallel;z,z')$ is the Green's function\cite{keller} for the
equation 
\bq
\tensor{L}(\omega,\vec{q}_\parallel)\cdot\tensor{G}(\omega,\vec{q}_\parallel;z,z')=\delta(z-z') .
\label{m13}\enq
We can easily solve Eq.(\ref{m13}) up to a homogeneous solution of
itself with its explicit expression shown in Appendix A.
We notice that, in all
equations, from Eq.(\ref{m11})to Eq.(\ref{m13}), $z$ and $z'$ are confined to the region
\bq
-\frac{L_c}{2}\le z\le\frac{L_c}{2}
\label{m14}\enq
Moreover, the integral equation (\ref{m11}) exhibits the nonlocality not
only provided by the Green's functions, but also induced by the medium
polarization effect which is described by the conductivity tensor.
To solve this integral equation, we emphasize that it should be
solved with proper boundary conditions (BC's). This is because 
neither the homogeneous solution
$\vec{E}^{(c)}(\omega,\vec{q}_\parallel,z)$ nor the Green's functions are
solved with respect to the correct BC's for the SMC. Therefore,
following classical electrodynamics, at
$z=\pm\frac{L_c}{2}$ we should have the BC 
\bqa
 &A_1+B_1=A_c e^{-i q_{\perp,c}\frac{L_c}{2}}+
B_c e^{i q_{\perp,c}\frac{L_c}{2}} \cr
& \hphantom{A_1+B_1= \ }
-\frac{4\pi i \omega}{c^2}\sum_{\nu'=x,z}\sum_{\nu''=x,z}
\int^{\Lambda/2}_{-\Lambda/2}dz'\int^{\Lambda/2}_{-\Lambda/2}dz''
G_{x,\nu'}\Bigl(\omega,\vec{q}_\parallel;-\frac{L_c}{2},z'\Bigr) \cr 
& \hphantom{A_1+B_1= \ }
\times\sigma_{\nu',\nu''}
(\omega,\vec{q}_\parallel;z',z'')E_{\nu''}(\omega,\vec{q}_\parallel;z'')
\label{m15}\enqa
\bqa
& \frac{\varepsilon_1}{q_{\perp,1}}(A_1-B_1)=\frac{\varepsilon_c}{q_{\perp,c}}
(A_c e^{-i q_{\perp,c}\frac{L_c}{2}}-
B_c e^{i q_{\perp,c}\frac{L_c}{2}})  \cr
& \hphantom{A_1+B_1= \ }
+\frac{\varepsilon_c}{q_{\perp,c}}
\frac{4\pi i\omega}{c^2}\sum_{\nu'=x,z}\sum_{\nu''=x,z}
\int^{\Lambda/2}_{-\Lambda/2}dz'\int^{\Lambda/2}_{-\Lambda/2}dz''
G_{x,\nu'}\Bigl(\omega,\vec{q}_\parallel;-\frac{L_c}{2},z'\Bigr) \cr 
& \hphantom{A_1+B_1= \ }
\times\sigma_{\nu',\nu''}
(\omega,\vec{q}_\parallel;z',z'')E_{\nu''}(\omega,\vec{q}_\parallel;z'')
\label{m16}\enqa
\bqa
& A_3+B_3=A_c e^{i q_{\perp,c}\frac{L_c}{2}}+
B_c e^{-i q_{\perp,c}\frac{L_c}{2}}  \cr 
& \hphantom{A_3+B_3=  }
-\frac{4\pi i \omega}{c^2}\sum_{\nu'=x,z}\sum_{\nu''=x,z}
\int^{\Lambda/2}_{-\Lambda/2}dz'\int^{\Lambda/2}_{-\Lambda/2}dz''
G_{x,\nu'}\Bigl(\omega,\vec{q}_\parallel;\frac{L_c}{2},z'\Bigr) \cr 
& \hphantom{A_3+B_3=  }
\times\sigma_{\nu',\nu''}
(\omega,\vec{q}_\parallel;z',z'')E_{\nu''}(\omega,\vec{q}_\parallel;z'')
\label{m17}
\enqa
\bqa 
&\frac{\varepsilon_1}{q_\perp,1}(A_3-B_3)=\frac{\varepsilon_c}{q_{\perp,c}}
(A_c e^{i q_{\perp,c}\frac{L_c}{2}}-
B_c e^{-i q_{\perp,c}\frac{L_c}{2}}) \cr
& \hphantom{\frac{\varepsilon_1}{q_\perp,1}(A_3-B_3)= \ }
-\frac{\varepsilon_c}{q_{\perp,c}}
\frac{4\pi i\omega}{c^2}\sum_{\nu'=x,z}\sum_{\nu''=x,z}
\int^{\Lambda/2}_{-\Lambda/2}dz'\int^{\Lambda/2}_{-\Lambda/2}dz''
G_{x,\nu'}\Bigl(\omega,\vec{q}_\parallel;\frac{L_c}{2},z'\Bigr)  \cr
& \hphantom{\frac{\varepsilon_1}{q_\perp,1}(A_3-B_3)= \ }
\times\sigma_{\nu',\nu''}
(\omega,\vec{q}_\parallel;z',z'')E_{\nu''}(\omega,\vec{q}_\parallel;z'')
\label{m18}
\enqa
for fixed $\omega$ and $\vec{q}_\parallel$, since $A_0$ is the input while
$B_4$ is usually taken to be zero. Here we have altogether eight independent
constants $B_0,A_1,B_1,A_c,B_c,A_3,B_3$ and $A_4$, and a
two-component electric field function $E_x(\omega,\vec{q}_\parallel,z)$ and 
$E_z(\omega,\vec{q}_\parallel,z)$ as unknown variables (functions), in
which $E_x$ and $E_z$ are defined  only in the SMC (including the QW). We
stress that they are mutually coupled to each other through Eqs.
(\ref{m6}), (\ref{m8}),(\ref{m15})-(\ref{m18}) and the integral equation Eq.(\ref{m11}) (referring also to Eqs. (\ref{m12}) and (\ref{A-1})-(\ref{A-3})).  Eqs. (\ref{m6}) and (\ref{m8}) are $2\times2$ matrix equations, the above equations are exactly eight
mutually coupled  algebraic equations and a two-component
integral equation, which solves self-consistently the whole DBR--SMC(QW)--DBR system as long as we have the detailed expression for the
conductivity tensor. Such a description \cite{liu} constitutes our
mathematical framework for studying semiclassically the excitonic polaritons in the QW embedded in a SMC . The pair of DBR's is
coherently correlated with the SMC and plays the role of mode
selection. Thus, the existing modes in the SMC not only
dynamically couple to the semiconductor medium in the QW, but also
coherently couple to the DBR's. This is the physical meaning
of such a description.
\section{ Polarization of the Semiconductor Medium  in a Quantum Well}
 In this section we  concentrate on the polarization of the semiconductor medium  in a quantum well. As mentioned in the introduction, the loss of displacement invariance in the growth direction of the QW and whole microstructure makes the dielectric function nonlocal. The non-locality of the medium polarization is closely related to the wave functions, and these functions are closely related to the band structure of the semiconductors. Thus,  first of all, we should give a brief description of the band structure of the semiconductor microstructures. It is well known that the electronic states in  a microstructure such as  a QW can be satisfactorily described by the envelope function approximation. However, in many publications, this approximation is  often an oversimplified version, i.e. the electron and holes are described as free particles with simple effective masses. Such a simplification usually works well in most cases, especially for conduction band electrons. However, this is not true for valence bands, i.e. for holes.  The top of the valence bands of the semiconductors with $T_d$ symmetry belongs to the  $\Gamma_8$ irreducible representation, which is 4-fold degenerate. In bulk material, the valence states apart from the $\Gamma$ point can be divided into heavy and light holes,  both of which exhibit strong anisotropy and non-parabolicity. For the microstructure case such as in  a QW, the situation is more complex. Generally speaking, any valence wavefunction can only be described by a four-component spinor, the basis of this spinor i.e. the Bloch  cell periodic function (with spin) satisfies certain space transformation relations, and more importantly, for different components the envelope functions  have different space rotational transformation properties. This invalidates the  simple free particle picture.   Although the problem of hole subband structure and related exciton's as well as the optical transitions  have been treated by several authors\cite{zhubf1}\cite{zhubf2}\cite{andreani2}, the complete analysis based on the exact non-local dielectric function and four-spinor nature of the wave function is still absent. We will devote the following section to this problem.\par
For the usual III-V compound semiconductors, at the conduction band bottom $\Gamma_6$, the Bloch cell periodic function  $u^{(c)}_\alpha(\vec{r}\,)$ is a spin doublet with index $\alpha=\frac{1}{2},-\frac{1}{2}$, while at the top of the valence band $\Gamma_8$ the cell
function  $u^{(v)}_{\alpha'}(\vec{r}\,)$ can be
regarded as a four component spinor with $\alpha'=\frac{3}{2},\frac{1}{2},
-\frac{1}{2},-\frac{3}{2}$, the HH subband corresponding to
$\alpha'=\pm\frac{3}{2}$, and the LH subband
 to $\alpha'=\pm\frac{1}{2}$. The plus and minus
signs refer to  time reversal (Kramer's)degenerate states. In the
effective mass approximation,  the non-interacting Hamiltonian for the envelope functions written in the first quantization representation has the form 
\bq
 H^{(0)}_{\alpha,\alpha';\beta,\beta'}=H^e_{\alpha,\beta}
\delta_{\alpha',\beta'}+\delta_{\alpha,\beta}H^h_{\alpha',\beta'}
\label{m19} 
\enq
\bq H^e_{\alpha,\beta}=\delta_{\alpha,\beta}\Bigl[\frac{1}{2m^e_\parallel}
\Bigl(\frac{\hbar}{i}\frac{\partial}{\partial\vec{r}_{\parallel,e}}\Bigr)^2+ 
\frac{1}{2m^e_\perp}\Bigl(\frac{\hbar}{i}\frac{\partial}{\partial z_e}\Bigr)^2+Eg-\mu+V_e
(\vec{r}_{\parallel,e},z_e)\Bigr]\label{m20}
\enq
\bq
 H^h_{\alpha',\beta'}=\frac{1}{2m}\pmatrix
{P_1, & Q, & R, & 0 \cr  Q^\ast, & P_2, & 0, & R \cr 
R^\ast, & 0, & P_2, & -Q \cr 
0, & R^\ast, & Q,& P_1} +\delta_{\alpha',\beta'}(\mu+V^{(\alpha')}_h
(\vec{r}_{\parallel,h},z_h))
\label{m21}
\enq
where 
\bqa
& P_1=(\gamma_1+\gamma_2)\Bigl(\frac{\hbar}{i}
\frac{\partial}{\partial\vec{r}_\parallel}\Bigr)^2+
(\gamma_1-2\gamma_2)\Bigl(\frac{\hbar}{i}
\frac{\partial}{\partial z_h}\Bigr)^2  \cr 
& P_2=(\gamma_1-\gamma_2)\Bigl(\frac{\hbar}{i}
\frac{\partial}{\partial\vec{r}_\parallel}\Bigr)^2+
(\gamma_1+2\gamma_2)\Bigl(\frac{\hbar}{i}
\frac{\partial}{\partial z_h}\Bigr)^2\cr 
& Q=-i2\sqrt3\gamma_3\Bigl(\frac{\hbar}{i}
\frac{\partial}{\partial z_h}\Bigr)
\Bigl(\Bigl(\frac{\hbar}{i}\frac{\partial}{\partial x_h}\Bigr)-i
\Bigl(\frac{\hbar}{i}
\frac{\partial}{\partial y_h}\Bigr)\Bigr)\cr 
& R=\sqrt3\Bigl[\gamma_2\Bigl(\Bigl(\frac{\hbar}{i}
\frac{\partial}{\partial x_h}\Bigr)^2-
\Bigl(\frac{\hbar}{i}\frac{\partial}{\partial y_h}\Bigr)^2\Bigr)-2i\gamma_3
\Bigl(\frac{\hbar}{i}\frac{\partial}{\partial x_h}\Bigr)
\Bigl(\frac{\hbar}{i}\frac{\partial}{\partial y_h}\Bigr)\Bigr]
\label{m22}
\enqa
In Eqs. (\ref{m20})-(\ref{m22}), the conduction electron coordinate is giving by 
$\vec{r}_e=(\vec{r}_{\parallel,e},z_e)$ with
$\vec{r}_{\parallel,e}=x_e\vec{e}_x+y_e\vec{e}_y$, and the valence hole
coordinate by $\vec{r}_h=(\vec{r}_{\parallel,h},z_h)$ with 
$\vec{r}_{\parallel,h}=x_h\vec{e}_x+y_h\vec{e}_y$; $E_g$ is the gap
which separates the conduction band and the valence band, $\mu$ 
the chemical potential, $m$ the physical electron mass, and
$m^e_\perp$ and $m^e_\parallel$ are effective masses of the
conduction electrons corresponding to motion
perpendicular or parallel to the $z$-axis; $V^e$ and $V^h$ are the
confinement potentials which form the barriers for the QW. For the
DBR--SMC(QW)--DBR system discussed in this paper, $V_e$ and $V_h$ are
actually $\vec{r}_{\parallel,e}$ and $\vec{r}_{\parallel,h}$
independent. Eq(\ref{m21}) with (\ref{m22}) is the well known Luttinger Hamiltonian\cite{Luttinger} which is written down in a form with a specific choice of the
coordinates. In these two equations, $Q$ and $R$ describe the
hybridization between the HH  and LH
subbands, while  $\gamma_1,\gamma_2$ and $\gamma_3$ are band structure parameters.
We notice that the envelope function description adopted in this paper
is valid around the $\Gamma$ point. We denote
$\varphi^{(b,\lambda)}_{s,\alpha}(\vec{r}\,)$ as the envelope function
which can be solved from the eigenvalue equation for the Hamiltonian
equations (\ref{m20}) and (\ref{m21}) with appropriate boundary conditions in connection
with the confinement barriers. Its corresponding energy eigenvalue
is denoted by $\varepsilon^{(b,\lambda)}_s$. In this notation, $b=c$
refers to the conduction band with $\lambda=\pm\frac{1}{2}$
corresponding to a spin doublet; $b=v$ refers to
the valence band with indices $\lambda=\pm\frac{1}{2}$,
$\pm\frac{3}{2}$ which describe valence band branches  resulting from diagonalization  of the hybridized Hamiltonian Eq.~(\ref{m21}). Notice that each of the eigenfunctions (envelope functions) corresponding to these branches is a four component spinor. We again name the envelope functions with $\lambda=\pm\frac{3}{2}$ as the HH branch and that with $\lambda=\pm\frac{1}{2}$
as the LH branch, according to the properties that the
$\lambda$-spinor has the dominant component as $\alpha=\lambda$.
Moreover, $s$ is the quantum number depending on the confinement
potential. It is discrete along the direction of confinement but
continuous in the extended directions.
\subsection{ Semiconductor Medium Polarization Induced by Virtual
Electron-Hole Pairs} 
It is known that the conductivity tensor is connected to the medium's
polarization tensor by
\bq
\tensor\sigma(\omega;\vec{r},\vec{r}\,')=
\frac{i}{\omega}\tensor{\pi}_r(\omega;\vec{r},\vec{r}\,)
\label{m23}
\enq
Intuitively, during light propagation, the dominant contribution to
the medium's polarization should be the virtual excitations of
electron-hole pairs. To clarify certain conceptual problems we
shall discuss the ideal situation: the intrinsic state of
the QW at low temperature, i.e., the valence band is almost fully filled
while the conduction band is almost completely empty. The virtual pairs thus consists
of  conduction electrons and valence holes. In such a
case, the Coulomb interaction should induce a series of exciton
states distributed in the semiconductor gap. These are bound states
formed by conduction electrons and valence holes. These virtual bounded
$e-h$ pairs should also contribute to the medium's  polarization even at zero
temperature for an intrinsic QW. 
Based upon such an understanding, the polarization can be
calculated straightforwardly as outlined in Appendix B.  We notice
that there are no requirements for the spatial
translational invariance, so that it can be applied to any sort of
QW. This also has the advantage that when applied to the planar
QW, the in-plane center of mass momentum for the excitons can be explicitly
treated, which is a non-trivial property for the excitonic polaritons. 
Now, the derived expression has the form 
\bqa
\tensor{\pi}_r(\omega;\vec{r},\vec{r}\,')=&\Bigl(\frac{e}{m}\Bigr)^2
\sum_\alpha\sum_{\alpha'}\sum_\beta\sum_{\beta'}
\langle v,\alpha'|\vec{p}|c,\alpha\rangle
\langle c,\beta|\vec{p}|v,\alpha\rangle\cdot \cr 
& \sum_{\lambda,s}\sum_{\lambda',s'}\sum_n\Bigl\{
\frac{\psi^{(\lambda,s;\lambda',s';n)}_{\alpha\alpha'}
(\vec{r},\vec{r}\,)\psi^{\ast(\lambda,s;\lambda',s';n)}_{\beta\beta'}
(\vec{r}',\vec{r}')}{\omega+i\eta-E^{(\lambda,s;\lambda',s';n)}_n} \cr 
& -\frac{\psi^{(\lambda,s;\lambda',s';n)}_{\alpha\alpha'}
(\vec{r}',\vec{r}')\psi^{\ast(\lambda,s;\lambda',s';n)}_{\beta\beta'}
(\vec{r},\vec{r}\,)}{\omega+i\eta+E^{(\lambda,s;\lambda',s';n)}_n}\Bigr\},
\label{m24}
\enqa
in which
\bq
\psi^{(\lambda,s;\lambda',s';n)}_{\alpha\alpha'}(\vec{r},\vec{r}\,)=
\psi^{(\lambda,s;\lambda',s';n)}_{\alpha\alpha'}(\vec{r}_e,\vec{r}_h)
|_{\vec{r}_e=\vec{r}_h=\vec{r}},
\label{m25}
\enq
where $\psi^{(\lambda,s;\lambda',s';n)}_{\alpha\alpha'}(\vec{r}_e,\vec{r}_h)$
is the two body exciton-like wave function for the $n$-th eigenstate
which consists of  a conduction electron with quantum numbers
$\lambda,s$ in $(\vec{r}_e,\alpha)$ representation and a valence hole
with quantum numbers $\lambda'$ and $s'$ in the $(\vec{r}_h,\alpha')$
representation. It satisfies 
\bqa
& \sum_{\beta,\beta'}\Bigl(H^{(0)}_{\alpha,\alpha';\beta,\beta'}-
\delta_{\alpha,\beta}\delta_{\alpha',\beta'}\frac{e^2}
{\varepsilon_c|\vec{r}_e-\vec{r}_h|}\Bigr)
\psi^{(\lambda,s;\lambda',s';n)}_{\beta\beta'}(\vec{r}_e,\vec{r}_h) \cr 
& \qquad =E^{(\lambda,s;\lambda',s')}_n
\psi^{(\lambda,s;\lambda',s';n)}_{\alpha\alpha'}(\vec{r}_e,\vec{r}_h),
\label{m26}
\enqa
where the eigenvalue $E^{(\lambda,s;\lambda',s')}_n$ is exactly the energy term appearing in Eq.(\ref{m24}). In Eq.(\ref{m24}), we have also introduced the
dipole matrix element 
\bq
\langle c,\alpha|\vec{p}|v,\alpha'\rangle\equiv\frac{1}{\Omega}\int_\Omega
d^3 r\frac{\hbar}{2i}(u^{\ast(c)}_\alpha(\vec{r}\,)\nabla 
u^{(v)}_{\alpha'}(\vec{r}\,)-\nabla u^{\ast(c)}_\alpha(\vec{r}\,)\cdot
u^{(v)}_{\alpha'}(\vec{r}\,)). \label{m27}
\enq
where $\Omega$ is the volume of the crystal cell and the integration
is also confined within the cell. We  emphasize here that, for the
inter-band processes, the dipole matrix element is carried by the
Bloch cell functions but not by the envelope functions.
In Eq.(\ref{m24}) only the two-body wave functions with its arguments   $\vec{r}_e=\vec{r}_h=\vec{r}$  contribute to the spectral weight of polarization propagator, $\vec r$ is the point where  the electron meets the hole .  Moreover, the first term in the bracket of Eq.~(\ref{m24}) is its positive frequency part while the second
is the negative frequency part. Both parts are necessary for a boson-like
propagator. We notice that the $\vec{r}$ and $\vec{r}\,'$ change into
each other in the two corresponding spectral weights, for which we understand that the positive and negative frequency parts correspond to a sort of
forward and backward propagation, respectively. Only when a certain kind of  symmetry is present for the system under investigation will the two spectral weight functions be equal to each other. This is one of the main topics for the next sub-section.
\par
 We notice that  the extended eigenstates of Eq.(\ref{m26}) can also contribute to the spectral function. Since they are off-resonance from  our interested frequencies, their contribution has been included  in the background dielectric constants.
\subsection{ Summation over the Degenerate States Based on Space-Time Inversion Symmetries}
After performing a summation over the space-time reversal transformation
connected degenerate states, we show in this subsection that not only the positive and negative
frequency parts in the polarization propagator can be combined into
one term, but also the non-diagonal elements for the polarization
tensor can diminish. This will make the expression
 neat and further simplifies the calculation a great deal.
\par
We apply the formal discussion in the previous sub-section to the system we interested in - a planar QW embedded in a DBR-SMC-DBR system (see
Sec.~II). The general subband indices and quantum numbers
$\lambda,s;\lambda',s'$ and $n$  are now specified into the following symbols: $\vec{q}_\parallel;n_{ex},l;n,j_c;n',$ and $j_h$. Here $\vec{q}_\parallel$ is the 2D center of mass momentum for the exciton, which is a good
quantum number describing the translational invariance in the
$X$-$Y$ plane for the system; $n_{ex}$ and $l$ are the quantum
numbers for the eigenstate of the exciton describing the in-plane
relative motion for the virtual electron and hole pairs; $n_{ex}$ is
the major quantum number while $l$ is the angular quantum number; $n$ ($n'$) is the index of the discrete states due to the confinement in the
$z$-direction of the conduction electrons (valence holes);
$j_c=\pm\frac{1}{2}$ are the subband indices describing the spin
doublet for the conduction electrons while
$j_h=\pm\frac{1}{2},~\pm\frac{3}{2}$ are the subband
indices for the valence holes representing the total angular momentum induced by spin-orbit coupling.
\par
To solve $\psi^{(\vec{q}_\parallel;n_{ex},l;n,j_c;n',j_h)}_{\alpha,\alpha'}(\vec{r}_e,
\vec{r}_h)$ from Eq.(\ref{m26}) we need the complete set of eigenfunctions of the non-interacting Hamiltonian described by Eqs (\ref{m20})
and (\ref{m21}) for the conduction electrons and valence holes, respectively. These are
\bq
\frac{1}{\sqrt{L^2}} e^{i\vec{k}_{\parallel,e}\cdot\vec{r}_{\parallel,e}}
\varphi^{(c)}_n(z_e)\delta_{j_c,\alpha}
\label{m28}
\enq
for the conduction electrons and
\bq
\frac{1}{\sqrt{L^2}} e^{i\vec{k}_{\parallel,h}\cdot\vec{r}_{\parallel,h}}
\varphi^{(v)}_{n',j_h;\alpha'}(\vec{k}_{\parallel,h};z_h)
\label{m29}
\enq
for the valence holes with subband hybridization, where $L^2$ is the
extension for the QW in the $X$-$Y$ plane. In solving
Eqs. (\ref{m20}) and (\ref{m21}), there is an energy eigenvalue $\epsilon^c_n$
associated with $\varphi^{(c)}_n(z_e)$ for the $z$-direction
electronic motion and another eigenvalue
$\epsilon^{(v)}_{n',j_h}=\epsilon^{(v)}_{n',j_h}(\vec{k}_\parallel=0)$
associated with 
$\varphi^{(v)}_{n',j_h;\alpha'}(\vec{k}_{\parallel,h},z_h)$ which
could also be  interpreted as describing the $z$-direction hole motion.
Actually, for a hole in the valence band, its motion along the
$z$-direction described by the eigen wave function
$\varphi^{(v)}_{n',j_h;\alpha'}$ is coupled to its motion in the
$X$-$Y$ plane characterized by $\vec{k}_{\parallel,h}$. So
$\varphi^{(v)}_{n',j_h;\alpha'}$ depends not only on $z_h$ but also
on $\vec{k}_{\parallel,h}$ whilst $\epsilon^{(v)}_{n',j_h}$ also has the $\vec{k}_{\parallel,h}$
dependence. Such a subtlety is due to the hybridization term $Q$ and
$R$ in Eq.~(\ref{m21}) and ~(\ref{m22})
\par
We notice further that, in the QW region,
$V^{(\alpha')}_h(\vec{r}_{\parallel,h},z)$ is constant so that we
chooce it to be zero.  The $z$-axis parities of  the matrix elements of the
Luttinger Hamiltonian Eq.(\ref{m21})
change their signs alternately along each row and each column.
This interesting symmetry property makes the hole eigenfunction
for Eq.(\ref{m21}) have an amazing parity symmetry so that the first and 
third components $(\alpha'=\frac{3}{2},-\frac{1}{2})$ have
the same parity symmetry along the $z$-axis which is opposite to that of the
other two components $(\alpha'=\frac{1}{2},-\frac{3}{2})$.
Moreover, since the Luttinger parameters $\gamma_2$ and $\gamma_3$
are usually rather close to each other, we may assume that they are
equal, i. e. $\gamma_2=\gamma_3$. We further introduce a polar coordinate
$(k_{\parallel,h},\vartheta_h)$ for $\vec{k}_{\parallel,h}$ in the
$X$-$Y$ plane, $\vec{k}_{\parallel,h}=k_{\parallel,h}(\cos\vartheta_h
\vec{e}_x+\sin\vartheta_h\vec{e}_y)$. Following Eq.(\ref{m22}), we 
have $Q\sim e^{-i\vartheta_h}$ and $R\sim e^{-2i\vartheta_h}$. As a
result, the corresponding eigenfunction in Eq.(\ref{m29}) acquires the
following functional dependence
\bq
\varphi^{(v)}_{n',j_h;\alpha'}(\vec{k}_{\parallel,h},z_h)= e^{i(j_h-\alpha')
\vartheta_h}\varphi^{(v)}_{n',j_h;\alpha'}(k_{\parallel,h}z_h)
\label{m30}\enq
in which $\vartheta_h$ is separated from
$\varphi^{(v)}_{n',j_h;\alpha'}(k_{\parallel,h}z_h)$ and the product
$k_{\parallel,h}z_h$ is a single dimensionless argument.
\par
We then expand the two-body exciton-like wave function in terms of
the non-interacting electron and hole wave functions, Eqs. (\ref{m28}) and
(\ref{m29}) for a fixed set of good quantum numbers
$\vec{q}_\parallel;n_{ex},l;n,j_c;n',j_h$, and obtain  
\bqa
& \psi^{(\vec{q}_\parallel;n_{ex},l;n,j_c;n',j_h)}_{\alpha,\alpha'}
(\vec{r}_e,\vec{r}_h)
=\frac{1}{L^3}\sum_{\vec{k}_{\parallel,e}}\sum_{\vec{k}_{\parallel,h}}
\delta_{\vec{q}_\parallel,\vec{k}_{\parallel,e}+\vec{k}_{\parallel,h}}
 e^{i(\vec{k}_{\parallel,e}\cdot\vec{r}_{\parallel,e}+\vec{k}_{\parallel,h}
\cdot\vec{r}_{\parallel,h})}\cdot \cr 
& \qquad
f^{(n;n',j_h)}_{n_{ex},l}(\vec{k}_{\parallel,e},\vec{k}_{\parallel,h})
\cdot\varphi^{(c)}_n(z_e)\delta_{j_c,\alpha}
\varphi^{(v)}_{n',j_h;\alpha'}(\vec{k}_{\parallel,h},z_h)
\label{m31}
\enqa
The expansion coefficient
$f^{(n;n',j_h)}_{n_{ex},l}(\vec{k}_{\parallel,e},\vec{k}_{\parallel,h})$
can be understood as the 2D excitonic wave function for which
the center of mass degree of freedom has been separated. If we ignore
the hybridization terms $Q$ and $R$ in the hole Hamiltonian Eq.~(\ref{m21}),
and further ignore the $z_e$ and $z_h$ dependence for the Coulomb
interaction in Eq.(\ref{m26}) because the thickness of the QW is rather small,
we can show easily that 
$f^{(n;n',j_h)}_{n_{ex},l}(\vec{k}_{\parallel,e},\vec{k}_{\parallel,h})$
would be precisely the 2D hydrogen atom solution of Eq.(\ref{m26}) constrained by
$\vec{q}_\parallel=\vec{k}_{\parallel,e}+\vec{k}_{\parallel,h}$.
Since now we have the hybridization, the physical meaning of 
$f^{(n;n',j_h)}_{n_{ex},l}(\vec{k}_{\parallel,e},\vec{k}_{\parallel,h})$
is no longer so transparent, and the coefficient should be determined by substituting
itself into Eq.~(\ref{m26}).
Actually we need only the value of the wave function (\ref{m31}) at $\vec{r}_e=\vec{r}_h=\vec{r}$. Thus, we have
\bqa
\psi^{(\vec{q}_\parallel;n_{ex},l;n,j_c;n',j_h)}_{\alpha,\alpha'}
(\vec{r},\vec{r})=\frac{1}{\sqrt{L^2}} e^{i\vec{q}_\parallel\cdot
\vec{r}_\parallel}\delta_{j_c,\alpha}\frac{1}{L^2}
\sum_{\vec{k}_{\parallel,e}}\sum_{\vec{k}_{\parallel,h}}
\delta_{\vec{q}_\parallel,\vec{k}_{\parallel,e}+
\vec{k}_{\parallel,h}}\cdot \cr 
\qquad f^{(n;n',j_h)}_{n_{ex},l}(\vec{k}_{\parallel,e},\vec{k}_{\parallel,h})
\varphi^{(c)}_n(z)\varphi^{(v)}_{n',j_h;\alpha'}(\vec{k}_{\parallel,h},z)
\label{m32}
\enqa
Moreover, the GaAs system has an energy gap $\sim1.5$ eV. The wavelength of interest has an order of magnitude
$\sim10^3$\AA. On the other hand, the effective Bohr radius for the 2D-exciton
in GaAs has the order of magnitude $\sim10^2$\AA.  Therefore,
$\vec{k}_{\parallel,e}$ and $\vec{k}_{\parallel,h}$ will span a
region in  momentum space in which the 2D excitonic wave function
$f^{(n;n',j_h)}_{n_{ex,l}}(\vec{k}_{\parallel,e},\vec{k}_{\parallel,h})$
has non-negligible contributions, much bigger than that of
$\vec{q}_\parallel$. We may reasonably set $\vec{q}_\parallel$ to be $\sim0$
in the double summation for $\vec{k}_{\parallel,e}$ and
$\vec{k}_{\parallel,h}$, i.e., we may have approximately
$\vec{k}_{\parallel,h}\cong-\vec{k}_{\parallel,e}=\vec{k}_\parallel$
in Eq.(\ref{m32}) within the double summation. Then
\bqa
& \psi^{(\vec{q}_\parallel;n_{ex},l;n,j_c;n',j_h)}
(\vec{r},\vec{r}) \cr 
& \qquad
=\frac{1}{\sqrt{L^2}} e^{i\vec{q}_\parallel\cdot
\vec{r}_\parallel}\delta_{j_c,\alpha}\int\frac{d \vec k_\parallel}{(2\pi)^2}
f^{(n;n',j_h)}_{n_{ex},l}(\vec{k}_{\parallel})
\varphi^{(c)}_n(z)\varphi^{(v)}_{n',j_h;\alpha'}(\vec{k}_{\parallel},z).
\label{m33}
\enqa
Intuitively, the 2D exciton-like wave function should have the
expression 
\bq
f^{(n;n',j_h)}_{n_{ex},l}(\vec{k}_{\parallel})=
f^{(n;n',j_h)}_{n_{ex},l}(k_{\parallel}) e^{il\vartheta},
\label{m34}
\enq
where $(k_\parallel,\vartheta)$ are  the polar coordinates of
$\vec{k}_\parallel$. Then, by further utilizing Eq.\ref{m30} and
$\int^{2\pi}_0 d\vartheta\exp$
$i(l+j_h-\alpha')\vartheta=2\pi\delta_{l+j_h,\alpha}$, we obtain
\bqa
& \psi^{(\vec{q}_\parallel;n_{ex},l;n,j_c;n',j_h)}_{\alpha,\alpha'}
(\vec{r},\vec{r}) \cr 
& \qquad
=\frac{1}{\sqrt{L^2}} e^{i\vec{q}_\parallel\cdot
\vec{r}_\parallel}\delta_{j_c,\alpha}\delta_{j_h+l,\alpha'}
\int^\infty_0\frac{k_\parallel d k_\parallel}{2\pi}
f^{(n;n',j_h)}_{n_{ex},l}(k_{\parallel})
\varphi^{(c)}_n(z)\varphi^{(v)}_{n',j_h;\alpha'}(k_{\parallel}z).
\label{m35}
\enqa
We substitute expression (\ref{m35}) into Eq.(\ref{m24}) with its indices changed
from the general form into the present, and obtain
\bqa
&\pi_{r;\mu,\nu}(\omega;\vec{r},\vec{r}\,')= \frac{e^2}{m^2}\frac{1}{L^2}
\sum_{\vec{q}_\parallel}\sum_{n_{ex},l}\sum_{n,j_c}\sum_{n',j_h}\langle
v,j_h+l|p_\mu|c,j_c\rangle\langle c,j_c|p_\nu|v,j_h+l\rangle \cr 
&\qquad \int^\infty_0\frac{k_\parallel d k_\parallel}{2\pi}
\int^\infty_0\frac{k'_\parallel d k'_\parallel}{2\pi}
f^{(n;n',j_h)}_{n_{ex},l}(k_{\parallel})
f^{(n;n',j_h)}_{n_{ex},l}(k'_{\parallel}) \cr 
&\qquad \varphi^{(c)}_n(z)\varphi^{(c)}_n(z')
\varphi^{(v)}_{n',j_h;\alpha'=j_h+l}(k_{\parallel}z)\cdot
\varphi^{(v)}_{n',j_h;\alpha'=j_h+l}(k'_{\parallel}z') \cr 
& \qquad\cdot\Bigl( e^{i\vec{q}_\parallel\cdot(\vec{r}_\parallel-
\vec{r}'_\parallel)}
\frac{1}{\omega+i\eta-E^{(n;n',j_h)}_{n_{ex},l}(q_\parallel)}-
 e^{-i\vec{q}_\parallel\cdot(\vec{r}_\parallel-\vec{r}'_\parallel)}
\frac{1}{\omega+i\eta+E^{(n;n',j_h)}_{n_{ex},l}(q_\parallel)}\Bigr),
\label{m36}
\enqa
where
\bq
E^{(n;n',j_d)}_{n_{ex},l}(q_\parallel)=E_g+
\epsilon^{(n;n',j_h)}_{q_\parallel}+
\epsilon^{(n;n',j_h)}_{n_{ex,l}}+\epsilon^{(c)}_n+
\epsilon^{(v)}_{n',j_h}
\label{m37}
\enq
 depends on $q_\parallel=|\vec{q}_\parallel|$ only. In Eq.~(\ref{m37}),
$\epsilon^{(n;n',j_h)}_{q_\parallel}$ is the kinetic energy for the
center of mass of the 2D exciton and
$\epsilon^{(n;n',j_h)}_{n_{ex,l}}\le0$ is its binding energy. The
$\epsilon^{(c)}_n$ and $\epsilon^{(v)}_{n'}$ are associated with the
confinement potential for the conduction electrons and valence holes
which has been introduced in connection with Eqs. (\ref{m28}) and (\ref{m29}). As
mentioned above, due to the hybridization of the valence holes, it is difficult in practice to reduce $E^{(n;n',j_d)}_{n_{ex},l}(q_\parallel)$ into such a neat form as Eq.(\ref{m37}) with conventional interpretations.  The function $f^{(n;n',j_h)}_{n_{ex},l}(k_{\parallel})$  used in Eq. (\ref{m36}) is essentially  the eigenfunction of the eigenvalue problem of Eq.(\ref{m26}) which can be solved either approximately or numerically. In practice, a variation solution was used. \par
 Besides the dynamical calculations, we have various
symmetry properties for the system which are exact. Due to the space-time
reversal symmetry, $E^{(n;n',j_d)}_{n_{ex},l}(q_\parallel)$ and
$f^{(n;n',j_d)}_{n_{ex},l}(k_\parallel)$ are degenerate with respect
to $\pm l$, $j_c=\pm\frac{1}{2}$, $j_h=\pm\frac{1}{2}$
and $j_h=\pm\frac{3}{2}$. We have also the time reversal
symmetry property for the hole wave function 
\bq
\varphi^{(v)}_{n',j_h;\alpha'}(z)=\varphi^{(v)}_{n',-j_h;-\alpha'}(z).
\label{m38}
\enq
Furthermore,  all the $f^{(n;n',j_h)}_{n_{ex},l}(k_\parallel)$,
$\varphi^{(c)}_n(z)$ and $\varphi^{(v)}_{n,j_d;\alpha'}(k_\parallel
z)$ are real functions. On the other hand, it is known that there is only one independent matrix element for $\langle
v,\alpha'|p_\mu|c,\alpha\rangle$, i.e.
$p=\langle X|p_x|s\rangle=\langle Y|p_y|s\rangle=\langle Z|p_z|s\rangle$ where $\langle X|,~\langle Y|$ and $\langle Z|$ are the orbital part of the Bloch cell functions for the hole band with $l=1$ and $|s\rangle$ is the orbital part of the Bloch cell functions for the conduction band. By making use of the Kramer's degeneracy properties and the Clebsch-Gordon coefficients for the matrix elements 
$\langle v,\alpha'|p_\mu|c,\alpha\rangle$, we can sum over the plus and minus
values of $l,j_z$ and $j_h$ but keep the absolute values  of $l,~j_c,~j_h$
and the corresponding energy eigenvalue fixed. After a lengthy calculation, the conductivity tensor associated with Eqs. (\ref{m23}) and (\ref{m36}) can be derived in a diagonal form as
\bq
\sigma_{\mu\nu}(\omega;\vec{r},\vec{r}\,')=\frac{1}{L^2}
\sum_{\vec{q}_\parallel}
 e^{i\vec{q}_\parallel\cdot(\vec{r}_\parallel-\vec{r}\,'_\parallel)}
\sigma_{\mu\nu}(\omega;\vec{q}_\parallel;z,z')
\label{m39}
\enq
with 
\bqa
\sigma_{\mu\nu}(\omega;\vec{q}_\parallel;z,z')=& \frac{i}{\omega}
\frac{e^2p^2}{m^2}
\delta_{\mu,\nu}\sum_n\sum_{n',j_h}\sum_{n_{ex},l}\sum_{l_h}
  \cr
& P^{(n;n',j_h)}_{n_{ex},l}(\omega,q_\parallel)\cdot
\phi^{(n;n',j_h)}_{n_{ex},l,l_h}(z)\eta^{(j_h;l,l_h)}_{\mu}
\phi^{\ast(n;n',j_h)}_{n_{ex},l,l_h}(z'),
\label{m40}
\enqa
in which
\bq
\phi^{(n;n',j_h)}_{n_{ex},l,l_h}(z')=\int^\infty_0\frac{k_\parallel dk_\parallel}{2\pi}f^{(n;n',j_h)}_{n_{ex},l}(k_\parallel)\varphi^{(c)}_n(z)
\varphi^{(v)}_{n',j_h;\alpha'=j_h+l_h}(k_\parallel z)
\label{m41}
\enq
\bq
P^{(n;n',j_h)}_{n_{ex},l}=\frac{2E^{(n;n',j_h)}_{n_{ex},l}(q_\parallel)}
{(\omega+i\gamma^{(n;n',j_h)}_{n_{ex},l})^2-[E^{(n;n',j_h)}_{n_{ex},l}
(q_\parallel)]^2}.
\label{m42}
\enq
In Eqs. (\ref{m40}),(\ref{m41}) and (\ref{m42}), $j_h$ takes the value of
$\frac{3}{2}$ and $\frac{1}{2}$ for HH and LH bands,
respectively, while $l$ is equal to $0$ and 1, $l_h=0$ if $l=0$, and$l_h$ should be summed over $\pm1$ if $l=1$. This ``new'' quantum
number $l_h$ results from the summation over the 
degenerate states due to Kramer's and space inversion symmetries. The term $\eta^{(j_h;l,l_h)}_\mu$ comes from the Clebsch-Gordon
coefficients for the dipole matrix elements, with
\bqa
& \eta^{(3/2,0,0)}_\mu=\eta^{(1/2,1,1)}_\mu=\pmatrix {1\cr  0 } \cr 
& \eta^{(1/2,0,0)}_\mu=\eta^{(3/2,1,1)}_\mu=\eta^{(1/2,1,-1)}_\mu=
\pmatrix {\frac{1}{3}\cr \frac{4}{3}} \cr 
& \eta^{(3/2,1,1)}_\mu=0.
\label{m43}
\enqa
We stress that if we ignore the hybridization between the HH
and LH subbands, then only $l=0$ exciton states
contribute to the spectral weight function. Meanwhile,
$\varphi^{(v)}_{n',j_h;\alpha'=j_h}(k_\parallel z)$ becomes 
$\varphi^{(v)}_{n',j_h;\alpha'=j_h}(z)$ which is independent of $k_\parallel$. Therefore, the hole subband hybridization induced
excitonic polaritons are characterized by the angular quantum number
 $l\ne 0$. We stress further that, in Eqs. (\ref{m40}) and (\ref{m42}), we replace
$\omega+i\eta$ by $\omega+i\gamma^{(n;n',j_h)}_{n_{ex},l}$ in which 
$\gamma^{(n;n',j_h)}_{n_{ex},l}$ are the only phenomenological
parameters introduced in this approach for describing the width of the
exciton level.
\par
\section{ Discrete Representation of Electric
Field in an SMC}
In the  last section, we obtained the  polarization of a medium
induced by the Coulomb interacting virtual electron-hole pairs in an
intrinsic semiconductor QW at zero temperature. We notice that the
$z,z'$ dependence for the conductivity tensor Eq.(\ref{m40}) is separated
into a bi-product of two $\phi$-functions. Thus, the kernel of the integral in  equation (\ref{m11}) is separable. Following Ref.\cite{liu} , we can easily transform the integral equation (\ref{m11})
for the electric field in an SMC into an algebraic equation which could
further simplify the calculations. In particular, as
$P^{(n;n',j_h)}_{n_{ex},l}(\omega,q_\parallel)$ has a resonant pole, only
very  few discrete components for the electric field are needed, and possibly only one may be sufficient. \par 
Introducing
\bq
\vec{F}^{(n;n',j_h)}_{n_{ex},l;l_h}(\omega,\vec{q}_\parallel)=
\int^{\Lambda/2}_{-\Lambda/2}dz\phi^{\ast(n;n',j_h)}_{n_{ex},l;l_h}(z)
\vec{E}(\omega,\vec{q}_\parallel;z),
\label{m44}
\enq
we may express the electric field in terms of
$\vec{E}^{(n;n',j_h)}_{n_{ex},l;l_h}(\omega,\vec{q}_\parallel)$ as
\bqa
 &\vec{E}(\omega,\vec{q}_\parallel,z)=
\vec{E}^{(c)}(\omega,\vec{q}_\parallel,z)+\frac{4\pi e^2p^2}{m^2c^2}\times
  \cr
 &\phantom{\vec{E}(\omega,\vec{q}_\parallel,z)=}\sum_{n,n',j_h}\sum_{n_{ex},l,l_h}
\int^{\Lambda/2}_{-\Lambda/2}dz'
P^{(n;n',j_h)}_{n_{ex},l}(\omega,\vec{q}_\parallel)
\tensor{G}(\omega,\vec{q}_\parallel,z,z')
 \cr 
 &\phantom{\vec{E}(\omega,\vec{q}_\parallel,z)=}\phi^{(n;n',j_h)}_{n_{ex},l;l_h}(z)\cdot
\tensor{\eta}^{(j_h,l,l_h)}\cdot
\vec{F}^{(n;n',j_h)}_{n_{ex},l;l_h}(\omega,\vec{q}_\parallel)
\label{m45}
\enqa
with
\bq
\tensor{\eta}^{(j_h,l,l_h)}=\delta_{\mu\nu}\eta^{(j_h,l,l_h)}_\mu
\label{m46}
\enq
We emphasize that  the $\vec{F}$-functions introduced by Eq.(\ref{m44}) depend
on the electric field defined only in the region of 
$[-\frac{\Lambda}{2},~\frac{\Lambda}{2}]$, but the
corresponding electric field  Eq.(\ref{m45}) is
meaningful for the whole region of  
$[-\frac{L_c}{2},~\frac{L_c}{2}]$. Utilizing the above two
equations, we can express Eq.(\ref{m11}) as well as the boundary condition
Eqs. (\ref{m15})-(\ref{m18}) in terms of $\vec{F}$-functions.  We perform the operation $\int^{\Lambda/2}_{-\Lambda/2}d
z\phi^{\ast(n;n',j_h)}_{n_{ex},l;l_h}(z)$ on both sides of Eq.~(\ref{m11}), and introduce $\vec{f}(\omega,\vec{q}_\parallel)$ and $\tensor{G}(\omega,\vec{q}_\parallel)$:
\bq
\vec{f}\,^{(n;n',j_h)}_{n_{ex},l;l_h}(\omega,\vec{q}_\parallel)=
\int^{\Lambda/2}_{-\Lambda/2}dz\phi^{\ast(n;n',j_h)}_{n_{ex},l;l_h}(z)
\vec{E}^{(c)}(\omega,\vec{q}_\parallel;z)
\label{m47} \enq
\bqa 
 &\tensor{G}^{(n;n',j_h;\;n'';n''',j'_h)}_{n_{ex},l,l_h;~n'_{ex},l'l'_h}
(\omega,\vec{q}_\parallel)=\int^{\Lambda/2}_{-\Lambda/2}dz
\int^{\Lambda/2}_{-\Lambda/2}dz' \cr 
&\hphantom{\tensor{G}^{(n;n',j_h;n'';n''',j'_h)}_{n_{ex},
l,l_h;n'_{ex},l'l'_h}(\omega,\vec{q}_\parallel) =}
\phi^{\ast(n;n',j_h)}_{n_{ex},l,l_h}(z)
\tensor{G}(\omega;\vec{q}_\parallel;z,z')
\phi^{(n'';n''',j'_h)}_{n'_{ex},l',l'_h}(z'),
\label{m48}
\enqa
 from which we obtain
\bqa
&\vec{F}^{(n;n',j_h)}_{n_{ex},l;l_h}(\omega,\vec{q}_\parallel)= 
\vec{f}^{(n;n',j_h)}_{n_{ex},l;l_h}(\omega,\vec{q}_\parallel)+
\frac{4\pi e^2p^2}{m^2c^2} \times\cr 
& \phantom{\vec{F}^{(n;n',j_h)}_{n_{ex},l;l_h}(\omega,\vec{q}_\parallel)=}
\sum_{n'',n''',j'_h}\sum_{n'_{ex},l',l'_h}
\tensor{ G}^{(n;n',j_h;\;n'';n''',j'_h)}_{n_{ex},l,l_h;~n'_{ex},l'l'_h}
(\omega,\vec{q}_\parallel)P^{(n'';n''',j'_h)}_{n'_{ex},l'}(\omega,\vec{q}_\parallel)\cdot\cr
& \phantom{\vec{F}^{(n;n',j_h)}_{n_{ex},l;l_h}(\omega,\vec{q}_\parallel)=}
 \cdot \tensor{\eta}^{(j'_h,l',l'_h)}\cdot
\vec{F}^{(n'';n''',j'_h)}_{n'_{ex},l';l'_h}(\omega,\vec{q}_\parallel).
\label{m49}
\enqa
The boundary condition Eqs. (\ref{m18}) can also be
reformulated in terms of
$F^{(n;n',j_h)}_{n_{ex},l;l_h}$ as
\bqa
& A_1+B_1=A_c e^{iq_{\perp,c}L_c/2}+B_c e^{iq_{\perp,c}L_c/2}+
 \cr 
& \hphantom{A_1+B_1= \ }
\frac{4\pi e^2p^2}{m^2c^2}
\sum_{n,n',j_h}\sum_{n_{ex},l,l_h}\sum_\mu
\int^{\Lambda/2}_{-\Lambda/2}dz'
P^{(n;n',j_h)}_{n_{ex},l}(\omega,\vec{q}_\parallel)
G_{x,\mu}\Bigl(\omega,\vec{q}_\parallel;-\frac{L_c}{2},z'\Bigr) \cr 
& \hphantom{A_1+B_1= \ }
\phi^{(n;n',j_h)}_{n_{ex},l;l_h}(z')
\eta^{(j_h,l,l_h)}_\mu F^{(n;n',j_h)}_{n_{ex},l;l_h;\mu}(\omega,\vec{q}_\parallel)
\label{m50}
\enqa
\bqa
& \frac{\varepsilon_1}{q_{\perp,1}}(A_1-B_1)=\frac{\varepsilon_c}{q_{\perp,c}}
(A_c e^{-iq_{\perp,c}L_c/2}-B_c e^{iq_{\perp,c}L_c/2})-
 \cr 
& \hphantom{\frac{\varepsilon_1}{q_{\perp,1}}(A_1-B_1)= \ }
\frac{\varepsilon_c}{q_{\perp,c}}\frac{4\pi e^2p^2}{m^2c^2}
\sum_{n;n',j_h}\sum_{n_{ex},l,l_h}\sum_\mu
\int^{\Lambda/2}_{-\Lambda/2}dz'
P^{(n;n',j_h)}_{n_{ex},l}(\omega,\vec{q}_\parallel)
G_{x,\mu} \label{m51}\enqa
\bqa
& \hphantom{\frac{\varepsilon_1}{q_{\perp,1}}(A_1-B_1)= \ }
\Bigl(\omega,\vec{q}_\parallel;-\frac{L_c}{2},z'\Bigr)
\phi^{(n;n',j_h)}_{n_{ex},l;l_h}(z')
\eta^{(j_h,l,l_h)}_\mu F^{(n;n',j_h)}_{n_{ex},l;l_h;\mu}(\omega,\vec{q}_\parallel)
\label{m52}\enqa
\bqa 
& A_3+B_3=A_c e^{iq_{\perp,c}L_c/2}+B_c e^{-iq_{\perp,c}L_c/2}
 \cr 
& \hphantom{A_3+B_3= \ }
_\frac{4\pi e^2p^2}{m^2c^2}
\sum_{n,n',j_h}\sum_{n_{ex},l;l_h}\sum_\mu
\int^{\Lambda/2}_{-\Lambda/2}dz
P^{(n;n',j_h)}_{n_{ex},l}(\omega,\vec{q}_\parallel)
G_{x,\mu}\Bigl(\omega,\vec{q}_\parallel;\frac{L_c}{2},z'\Bigr) \cr 
& \hphantom{A_3+B_3= \ }
\phi^{(n;n',j_h)}_{n_{ex},l;l_h}(z')
\eta^{(j_h,l,l_h)}_\mu F^{(n;n',j_h)}_{n_{ex},l;l_h;\mu}(\omega,\vec{q}_\parallel)
\label{m53}\enqa
\bqa
& \frac{\varepsilon_1}{q_{\perp,1}}(A_3-B_3)=\frac{\varepsilon_c}{q_{\perp,c}}
(A_c e^{iq_{\perp,c}L_c/2}+B_c e^{-iq_{\perp,c}L_c/2})+
 \cr 
& \hphantom{\frac{\varepsilon_1}{q_{\perp,1}}(A_3-B_3)= \ }
\frac{\varepsilon_c}{q_{\perp,c}}\frac{4\pi e^2p^2}{m^2c^2}
\sum_{n;n',j_h}\sum_{n_{ex},l,l_h}\sum_\mu
\int^{\Lambda/2}_{-\Lambda/2}dz'
P^{(n;n',j_h)}_{n_{ex},l}(\omega,\vec{q}_\parallel)
G_{x,\mu} \cr 
& \hphantom{\frac{\varepsilon_1}{q_{\perp,1}}(A_1-B_1)= \ }
\Bigl(\omega,\vec{q}_\parallel;\frac{L_c}{2},z'\Bigr)
\phi^{(n;n',j_h)}_{n_{ex},l;l_h}(z')
\eta^{(j_h,l,l_h)}_\mu F^{(n;n',j_h)}_{n_{ex},l;l_h;\mu}
(\omega,\vec{q}_\parallel).
\label{m54}
\enqa
Then, in terms of the discrete representation for the electric field 
$\vec{F}^{(n;n',j_h)}_{n_{ex},l;l_h}(\omega,\vec{q}_\parallel)$, the
complete description of the DBR-SMC(QW)-DBR system is now given by Eqs.
(\ref{m51})-(\ref{m54}) and (\ref{m49}), referring also to Eqs.(\ref{m6}), (\ref{m8}),(\ref{m12}),(\ref{A-3}),(\ref{m44}), (\ref{m45})and (\ref{m55}). Note that in order to express $\vec{f}\,^{(n;n',j_h)}_{n_{ex},l;l_h}(\omega,\vec{q}_\parallel)$ in Eq. (\ref{m49}) in terms of $A_c$ and $B_c$  we need to introduce two more terms  $w\,^{(n,n',j_h}_{n_{ex},l;l_h}(\omega,\vec{q}_\parallel)$ and $w'\,^{(n,n',j_h}_{n_{ex},l;l_h}(\omega,\vec{q}_\parallel)$ \bqa
w\,^{(n,n',j_h}_{n_{ex},l;l_h}(\omega,\vec{q}_\parallel)=
\int^{\Lambda/2}_{-\Lambda/2}dz\,e^{i q_{\perp,c}z} \phi^{\ast(n;n',j_h)}_{n_{ex},l;l_h}(z)
\vec{E}^{(c)}(\omega,\vec{q}_\parallel;z)\cr
w'\,^{(n,n',j_h}_{n_{ex},l;l_h}(\omega,\vec{q}_\parallel)=
\int^{\Lambda/2}_{-\Lambda/2}dz\,e^{-i q_{\perp,c}z} \phi^{\ast(n;n',j_h)}_{n_{ex},l;l_h}(z)
\vec{E}^{(c)}(\omega,\vec{q}_\parallel;z)
\enqa 
according to Eq \ref{m12}. Up to now, it is clear that we do not
need the effective photon-exciton coupling constant any more in the
approach we have adopted.\par
We have now  established a complete set of equations which can give  us the  real space electric field distribution  within SMC with non-local dielectric response of QW excitons embedded in it, and thus enable us to obtain various semiclassical optical properties. Compared with previous theoretical work, our approach treats both the electro-magnetic field and the electron excitation exactly and does not depend on any artificial parameters. Moreover , certain effects not considered in previous studies such as violation of simple selection rules and the strange angle dependence of the reflection spectrum, can be easily calculated using this approach. We will report these results elsewhere. As an application, we analyze some interesting aspects of the exciton "selection rule" for a symmetric  microcavity. \par
 It is well known from ordinary exciton models that only the $S$ exciton can couple to the electromagnetic field, but due to the  four component nature of the hole subband  wave function and the existence of band mixing, the $p$ exciton can also couple to the  photon field and thus form polaritons in a microcavity. We call this  a hybridization-induced exciton-polariton, and analyze the  relevant parity propeties along the $z$ axis.\par
It can be seen that Eq.(\ref{m45}) in conjunction with expressions (\ref{m41}) and (\ref{m44}) shows explicitly how the non-locality for the conductivity behaves. Each term of the spectral weight function has the  form of a product of two $\phi$-functions with one of the two depending only on $z$ and the other on $z'$. Therefore, one of them is convoluted with the electric field and plays a role as a ``weight'' function while the other is convoluted with the Green's function as a ``source'' function, which strongly influences the effective coupling between light and the exciton. This makes the role of the parity symmetry along thr $z$-axis explicitly important.
Here we restrict ourself to the case of normal incidence and
symmetric DBR pairs only. As we have discussed in Sec.~II-B, $\varphi^{(c)}_n(z)$ and $\varphi^{(v)}_{n,j_h;\alpha'}(k_\parallel
z)$ can be classified by the parity symmetry (along $z$-axis).
In fact, the parity symmetry for the contributive component of
the hole spinor wave functions is controlled by its index $\alpha'=j_h+l_h$.
Moreover, for an SMC confined by a pair of symmetric DBR's, it is
known that $\vec{E}^{(c)}(\omega,\vec{q}_\parallel;z)$ as a function
of $z$ is also an eigenstate of the parity symmetry at the resonant
condition $L_c=\frac{n}{2}\lambda$, with $n=1,2,\cdots$ . By considering Eqs. (\ref{m11}), (\ref{A-3}), (\ref{m40}) and (\ref{m41}), we can
verify that the electric field $\vec{E}(\omega,\vec{q}_\parallel;z)$
will have the same even-oddness as that of
$\vec{E}^{(c)}(\omega,\vec{q}_\parallel;z)$. Following Eqs. (\ref{m45}), (\ref{m4})
and (\ref{m41}), the physics contributed by the QW depends strongly on the integral
\bq
I^{(n;n',j_h)}_{l_h}(\omega,\vec{q}_\parallel;k_\parallel)=
\int^{\Lambda/2}_{-\Lambda/2}dz\varphi^{(c)}_n(z)
\varphi^{(v)}_{n',j_h;\alpha'=j_h+l_h}(k_\parallel
z)E_x(\omega,\vec{q}_\parallel;z),
\label{m55}
\enq 
and we can now find interesting consequences for the polaritons.
If we set $\varphi^{(c)}_n(z)$ for the conduction electron to be always
in the $n=1$ state, it will not contribute to the parity
symmetry since it is an even function of $z$. For a symmetric SMC,  $E^{(c)}(\omega,q_\parallel;z)$ is even for
$L_c=\frac{\Lambda}{2}$ and  odd for $L_c=\lambda$. Hence  for an
SMC of $L_c=\frac{\Lambda}{2}$, the electric field is
even. Then, following Eq.(\ref{m35}), only polaritons with quantum number
$l=l_h=0$ and the index $\alpha'=j_h$ can survive in the SMC for both
$j_h=\frac{1}{2}$  and $j_h=\frac{3}{2}$. This is because 
$\varphi^{(v)}_{n'=1,j_h;\alpha'=j_h}(k_\parallel z)$ is an even
function of $z$. The hybridization induced polaritons with quantum number $l=1$ cannot be observed since
$\varphi^{(v)}_{n'=1,j_h;\alpha'=j_h\pm1}(k_\parallel z)$ are odd
functions of $z$ which will make
$I^{(1;1,j_h)}_{{l_h=\pm1}(\omega,\vec{q}_\parallel;k_\parallel)}=0$. On the contrary, if we have an SMC of $L_c=\lambda$, the electric field $E_x(\omega,\vec{q}_\parallel;z)$ becomes an odd function of $z$, if  $\varphi^{(v)}_{n'=1,j_h;\alpha'=j_h}(k_\parallel z)$ is
still an even function, the polaritons with quantum number $n=n'=1$,
$l=0$ will be forbidden for the $\lambda$-SMC which will make 
$I^{(1;1,j_h)}_{l_h}(\omega,\vec{q}_\parallel;k_\parallel)=0$. In this case, only the hybridization induced polariton can survive in the SMC since $\varphi^{(v)}_{n'=1,j_h;\alpha'=j_h\pm l_h}(k_\parallel z)$
is odd. What we learned from the above discussion is that the
even-oddness of the electric field depends on the cavity resonance condition, while the parity of the components of the electron or hole wave function depends only on the QW. Then we can have an interplay of the even-odd symmetry for the integrand of Eq.(\ref{m44}). As a result the parity symmetry along the $z$-axis will provide a sort of selection rule for the forbidden polaritons in a symmetric DBR pair confined resonant SMC. This  could be helpful for finding the hybridization induced polaritons. Our calculation  shows the existence of HH subband dominated polaritons which have the quantum numbers $n$=1, $n'$=2, $j_h$=$\frac{3}{2}, n_{ex}=2, ~l=1$ with splitting value $\sim0.2$ mev. The details and other interesting results will be published elsewhere.
\par

\section{ Concluding remarks}
In conclusion, we have presented a self-consistent semiclassical approach for exciton-polaritons in  a QW embedded in an SMC. In this approach, the effect of the complex valence band structure  and the non-locality of the dielectric response of the exciton in the QW are carefully considered.  For 1HH excitons and normal incidence, our approach gives the same results  as those obtained previously. For complex cases such as high index excitons and oblique incidence we expect that it could predict some new phenomena. For example, we have shown  that a 2P exciton can also couple to a  photon mode and form  a polariton.  Moreover our analysis gives a "selection rule" for the formation of  exciton-polaritons in a symmetric SMC, which is essentially an interplay among the angular quantum numbers of excitons, the electron-hole subbands indices and the resonance conditions of the SMC. 
\acknowledgments
The authors are grateful to Drs. Jizhong Lou, Shaojin Qin and Bangfen Zhu for beneficial discussions and kind help. The authors acknowledge  support from the Chinese Academy of Sciences.\par

\begin{figure}
\caption{Schematics of the symmetric planar semiconductor microcavity}
\label{MC}
\end{figure}

\begin{figure}
\caption{The summed Coulomb interacting electron-hole bubble diagram series}
\label{bubble}
\end{figure}

\appendix
\section{ The Transfer Matrix Elements and Green's Functions.}
The transfer matrix elements can be calculated straightforwardly as\cite{yariv}
\bqa
& t^{ L}_{11}= e^{-i q_{\perp,1}{ L}_1}\Bigl[\cos (q_{\perp,2}L_2)-\frac{i}{2}\Bigl(\frac{\varepsilon_2q_{\perp,1}}{\varepsilon_1q_{\perp,2}}
+\frac{\varepsilon_1q_{\perp,2}}{\varepsilon_2q_{\perp,1}}\Bigr)\sin
q_{\perp,2}L_2\Bigr]
\cr
& t^{ L}_{12}= e^{i q_{\perp,1}{ L}_1}\Bigl[-\frac{i}{2}\Bigl(\frac{\varepsilon_2q_{\perp,1}}{\varepsilon_1q_{\perp,2}}
-\frac{\varepsilon_1q_{\perp,2}}{\varepsilon_2q_{\perp,1}}\Bigr)\sin
q_{\perp,2}L_2\Bigr]
\cr 
& t^{ L}_{21}= e^{-i q_{\perp,1}{ L}_1}\Bigl[
\frac{i}{2}\Bigl(\frac{\varepsilon_2q_{\perp,1}}{\varepsilon_1q_{\perp,2}}
-\frac{\varepsilon_1q_{\perp,2}}{\varepsilon_2q_{\perp,1}}\Bigr)\sin
q_{\perp,2}L_2\Bigr]
\cr 
& t^{ L}_{22}= e^{i q_{\perp,1}{ L}_1}\Bigl[\cos q_{\perp,2}L_2
-\frac{i}{2}\Bigl(\frac{\varepsilon_2q_{\perp,1}}{\varepsilon_1q_{\perp,2}}
+\frac{\varepsilon_1q_{\perp,2}}{\varepsilon_2q_{\perp,1}}\Bigr)\sin
q_{\perp,2}L_2\Bigr]
\label{A-1}
\enqa
and
\bqa
& t^{ R}_{11}=t^{ L}_{11}\,, \qquad & t^{ R}_{22}=t^{L}_{22},  \cr 
& t^{ R}_{12}=-t^{ L}_{21}\,, \qquad & t^{ R}_{21}=-t^{L}_{12}, 
\label{A-2}
\enqa
Moreover, the Green's functions in Eq.~(\ref{m13}) can also be easily
solved as\cite{keller}
\bqa
& G_{x,x}(\omega,q_\parallel;z,z')=-
\frac{i c^2}{2\omega^2\varepsilon_c(\omega)}
q_{\perp,c} e^{i q_{\perp,c}|z-z'|}
\cr 
& G_{x,z}(\omega,q_\parallel;z,z')=G_{z,x}(\omega,q_\parallel;z,z')=
\frac{i c^2}{2\omega^2\varepsilon_c(\omega)}q_\parallel Sgn(z-z')
 e^{i q_{\perp,c}|z-z'|}
\cr 
& G_{z,z}(\omega,q_\parallel;z,z')=\frac{c^2}
{\omega^2\varepsilon_c(\omega)}\delta(z-z')-
\frac{i c^2}{2\omega^2\varepsilon_c(\omega)}\frac{q^2_\parallel}{q_{\perp,c}}
 e^{i q_{\perp,c}|z-z'|}
\cr 
& G_{y,y}(\omega,q_\parallel;z,z')=-\frac{i}{2q_{\perp,c}}
 e^{i q_{\perp,c}|z-z'|}
\cr 
& G_{xy}=G_{yx}=G_{yz}=G_{zy}=0
\label{A-3}
\enqa

\section{Outline of the Derivation of the
Polarization Propagator}
We outline here the main steps for deriving Eq.(\ref{m24}) by
a summation over the Coulomb interaction ladder diagrams accommodated
in a bubble diagram consisting of a conduction electron and a
valence hole, as shown in Fig.~\ref{bubble}.
\par
We introduce the second quantized wave operator and current operator: 
\bq
\hat{\psi}(\vec{r}\,)=\sum_b\sum_\lambda\sum_s\sum_\alpha
\varphi^{(b,\lambda)}_{s,\alpha}(\vec{r}\,)u^{(b)}_\alpha
(\vec{r}\,)\hat{c}^{(b,\lambda)}_s
\label{B-1}
\enq
\bq
\hat{\vec{j}}(\vec{r}\,)=\frac{e\hbar}{2imc}(\hat{\psi}^\dag(\vec{r}\,)
\nabla\hat{\psi}^\dag(\vec{r}\,)-\nabla
\hat{\psi}^\dag(\vec{r}\,)\hat{\psi}(\vec{r}\,))
\label{b_2}
\enq
in which $\hat{c}^{(b,\lambda)}_s$ is the second quantized electron
annihilation operator for a state with band index $c,\lambda$ and
quantum number $s$. Note that we dropped the index $k$ of the cell-periodic functions of the Bloch function, in corresponding with the envelope function approximation . Using the linear response theory approximation  and the Mastubara representation, the polarization part of the electron system ${\pi}_r(\omega;\vec{r},\vec{r}\,')$ can be expressed as
 \bq
\tensor{\pi}_r(\omega;\vec{r},\vec{r}\,')=
-\int^\infty_0 d\tau e^{i\omega_n\tau} Tr
\{\hat{\rho}T_\tau\hat{\vec{j}}(\vec{r},\tau)\hat{\vec{j}}(\vec{r}\,',o)\}
|_{i\omega_n\to\omega+i\eta}
\label{B-6}
\enq
where $\hat{\rho}$ is the thermal density matrix, $ Tr$ the
trace operation, and $T_\tau$ the chronological operation along the imaginary time axis. Expression (\ref{B-6}) has the advantage that it can
be calculated systematically by applying the diagrammatical technique.
We notice first that each pair of electron and hole lines in the
upper series of diagrams in Fig.~\ref{bubble} contributes a  term 
\bq
\frac{1-n(\xi^{(c,\lambda)}_s)-n(\xi^{(v,\lambda')}_{\bar s'})}
{i\omega_n-\xi^{(c,\lambda)}_s-\xi^{(v,\lambda')}_{\bar s'}}
\label{B-7}
\enq
where $\bar{s}'$ describes the quantum numbers charge conjugated to 
$s'$. Each pair of electron and hole lines in the lower series of
diagrams in Fig.~\ref{bubble} with the  directions of the arrows reversed, 
contributes a term 
\bq
-\frac{1-n(\xi^{(c,\lambda)}_s)-n(\xi^{(v,\lambda')}_{\bar s'})}
{i\omega_n+\xi^{(c,\lambda)}_s+\xi^{(v,\lambda')}_{\bar s'}}
\label{B-8}
\enq
in which $\xi^{(c,\lambda)}_s=\varepsilon^{(c,\lambda)}_s+E_g-\mu$, $\xi^{(v,\lambda)}_s=\varepsilon^{(v,\lambda)}_s+\mu$ and n$(\xi)$ is
the Fermi distribution function for the conduction electrons and valence
holes. Moreover,
a dotted Coulomb line contributes a factor 
\bq
v^{\lambda,\lambda'';\lambda',\lambda'''}_{s',s'';s',s'''}=
\sum_{\alpha,\alpha'}\int d^3r d^3r'\varphi^{\ast(c,\lambda)}_{s,\alpha}
(\vec{r}\,)\varphi^{(c,\lambda'')}_{s'',\alpha}(\vec{r}\,)
\frac{1}{\varepsilon_c|\vec{r}-\vec{r}\,'|}
\varphi^{\ast(v,\lambda')}_{s',\alpha'}(\vec{r}\,)
\varphi^{(v,\lambda''')}_{s''',\alpha'''}(\vec{r}\,')
\label{B-4}
\enq
where $\varepsilon_c=\varepsilon_c(\omega)$ is the background dielectric constant for the medium.  Finally, each vertex at the end point of the bubble contributes a dipole matrix element $\langle c,\alpha|\vec{p}|v,\alpha'\rangle$ or $\langle v,\alpha'|\vec{p}|c,\alpha \rangle$. All the above expressions follow straightforwardly from the quantum many-body text-book with an additional consideration shown in (\ref{B-1}), i.e. our basis wave function is not the usual simple plane wave, but the envelope function associated with a more "microscopic" cell periodic function $u$.
\par
 It should also be noticed that, for an intrinsic semiconductor in the  low excitation limit, the conduction band is almost completely empty and the valence band almost completely full, so the two Fermi distribution functions in Eqs. (\ref{B-7}) and (\ref{B-8}) can be taken as zero. Furthermore, this guarantees that  only the ladder diagrams give any contribution\cite{mahan}. Through summation over the Coulomb interacting ladder diagrams accommodated in a bubble diagram which consit of a conduction electron and a valence hole  (shown in Fig.~\ref{bubble}), we can obtain the following expression for the polarization tensor, after a direct but lengthy calculation, 
\bqa
& \tensor{\pi}(i\omega_n;\vec{r},\vec{r}\,')=\Bigl(\frac{e}{m}\Bigr)^2
\sum_{\alpha,\alpha'}\sum_{\lambda,s}\sum_{\lambda',s'}\Bigl\{
\varphi^{(c,\lambda)}_{s,\alpha}(\vec{r}\,)
\varphi^{\ast(v,\lambda')}_{s',\alpha'}(\vec{r}\,)\langle v,\alpha'
|\vec{p}|c,\alpha\rangle   \cr 
&\hphantom{{\pi}(i\omega_n;\vec{r},\vec{r}\,')= \ }
\frac{1}{i\omega_n-\xi^{(c,\lambda)}_s-\xi^{(v,\lambda')}_{\bar s'}}
\Gamma^{\lambda,\lambda'}_{s,s'}(i\omega_n;\vec{r}\,') \cr 
&\hphantom{{\pi}(i\omega_n;\vec{r},\vec{r}\,')= \ }
-\varphi^{\ast(c,\lambda)}_{s,\alpha}(\vec{r}\,)
\varphi^{(v,\lambda')}_{s',\alpha'}(\vec{r}\,)\langle c,\alpha
|\vec{p}|v,\alpha'\rangle  \cr 
&\hphantom{{\pi}(i\omega_n;\vec{r},\vec{r}\,')= \ }
\frac{1}{i\omega_n+\xi^{(c,\lambda)}_s+\xi^{(v,\lambda')}_{\bar s'}}
\tilde{\Gamma}^{\lambda,\lambda'}_{s,s'}(i\omega_n;\vec{r}\,')\Bigr\},
\label{B-9} \cr 
\enqa 

\bqa 
& \Gamma^{\lambda,\lambda'}_{s,s'}
(i\omega_n;\vec{r}\,')=\sum_{\alpha'',\alpha'''}
\varphi^{\ast(c,\lambda)}_{s,\alpha''}(\vec{r}\,')
\varphi^{(v,\lambda')}_{s',\alpha'''}(\vec{r}\,')\langle c,\alpha''
|\vec{p}\,'|v,\alpha'''\rangle \cr 
&\hphantom{\Gamma^{\lambda,\lambda'}_{s,s'}(i\omega_n;\vec{r}\,')= \ }
+(-e^2)\sum_{\lambda'',s''}\sum_{\lambda''',s'''}
v^{\lambda,\lambda'';\lambda''',\lambda'}_{s,s'';s''',s'}\cdot
 \cr 
&\hphantom{\Gamma^{\lambda,\lambda'}_{s,s'}(i\omega_n;\vec{r}\,')= \ }
\frac{1}
{i\omega_n-\xi^{(c,\lambda'')}_{s''}-\xi^{(v,\lambda''')}_{\bar s'''}}
\Gamma^{(\lambda'',\lambda''')}_{s'',s'''}(i\omega_n;\vec{r}\,'),
\label{B-10} \cr
\enqa
\bqa 
& \tilde{\Gamma}^{\lambda,\lambda'}_{s,s'}
(i\omega_n;\vec{r}\,')=\sum_{\alpha'',\alpha'''}
\varphi^{(c,\lambda)}_{s,\alpha''}(\vec{r}\,')
\varphi^{\ast(v,\lambda')}_{\bar s',\alpha'''}(\vec{r}\,')\langle v,\alpha'''
|\vec{p}\,'|c,\alpha''\rangle \cr \
&\hphantom{\tilde{\Gamma}^{\lambda,\lambda'}_{s,s'}(i\omega_n;\vec{r}\,')= \ }
-(-e^2)\sum_{\lambda'',s''}\sum_{\lambda''',s'''}
v^{\lambda'',\lambda;\lambda',\lambda'''}_{s'',s;s',s'''} \cr \
&\hphantom{\Gamma^{\lambda,\lambda'}_{s,s'}(i\omega_n;\vec{r}\,')= \ }
\frac{1}
{i\omega_n+\xi^{(c,\lambda'')}_{s''}+\xi^{(v,\lambda''')}_{\bar s'''}}
\tilde{\Gamma}^{\lambda'',\lambda'''}_{s'',s'''}(i\omega_n;\vec{r}\,').
\label{B-11}
\enqa
We then introduce two auxiliary functions 
\bqa
P_{\alpha,\alpha'}(i\omega_n;\vec{\tilde{r}},\vec{\tilde{r}'};\vec{r}\,')=&
\sum_{\tilde{\lambda},\tilde{s}}\sum_{\tilde{\lambda}',\tilde{s}'}
\varphi^{(c,\tilde{\lambda})}_{\tilde{s},\alpha}(\vec{\tilde{r}}\,')
\varphi^{\ast(v,\tilde{\lambda}')}_{\bar{\tilde{s}}',\alpha'}
(\vec{\tilde{r}}\,') \cr 
& \frac{1}
{i\omega_n-\xi^{(c,\tilde{\lambda})}_{\tilde{s}}-
\xi^{(v,\tilde{\lambda}')}_{\bar{\tilde{s}'}}}
\Gamma^{(\tilde{\lambda},\tilde{\lambda}')}_{\tilde{s},\tilde{s}'}
(i\omega_n;\vec{r}\,')
\label{B-12}
\enqa
and
\bqa
\tilde{P}_{\alpha,\alpha'}(i\omega_n;\vec{\tilde{r}},\vec{\tilde{r}'};
\vec{r}\,')=&
\sum_{\tilde\lambda,\tilde s}\sum_{\tilde\lambda',\tilde s'}
\varphi^{\ast(c,\tilde\lambda)}_{\tilde s,\alpha}(\vec{\tilde{r}})
\varphi^{(v,\tilde\lambda')}_{\bar{\tilde s}',\alpha'}
(\vec{\tilde{r}}\,') \cr 
&\frac{1)}
{i\omega_n+\xi^{(c,\tilde\lambda)}_{\tilde s}+
\xi^{(v,\tilde\lambda')}_{\bar{\tilde{s}'}}}
\tilde{\Gamma}^{(\tilde\lambda,\tilde\lambda')}_{\tilde s,\tilde s'}
(i\omega_n;\vec{r}\,').
\label{B-13}
\enqa
We can express
$\Gamma^{(\lambda,\lambda)}_{s,s'}(i\omega_n,\vec{r}\,')$ and 
$\tilde{\Gamma}^{(\lambda,\lambda)}_{s,s'}(i\omega_n,\vec{r}\,')$ in
terms of the two auxiliary functions by utilizing the orthogonal
properties of $\varphi^{(c,\lambda)}_{s,\alpha}(\vec{r})$ and
$\varphi^{(v,\lambda')}_{s',\alpha'}(\vec{r})$, in a way  similar to an
inverse generalized Fourier transformation in the Hilbert space. We
have further the completeness relation for
$\varphi^{(c,\lambda)}_{s,\alpha}(\vec{r})$ and
$\varphi^{(v,\lambda')}_{s',\alpha'}(\vec{r}\,')$. With the aid of these
procedures, the polarization tensor can now be derived as
\bqa
\tensor{\pi}(i\omega_n;\vec{r},\vec{r}\,')=& \Bigl(\frac{e}{m}\Bigr)^2
\sum_{\alpha,\alpha'}\{\langle v,\alpha'|\vec{p}|c,\alpha\rangle
P_{\alpha,\alpha'}(i\omega_n;\vec{r},\vec{r};\vec{r}\,') \cr 
& -\langle c,\alpha|\vec{p}|v,\alpha'\rangle
\tilde{P}_{\alpha,\alpha'}(i\omega_n;\vec{r},\vec{r};\vec{r}\,')\}.
\label{B-14}
\enqa
Next, we apply the operator
$i\omega_n\delta_{\alpha,\alpha''}\delta_{\alpha',\alpha'''}-[H_0]_{
\alpha,\alpha';\alpha'',\alpha'''}$ to the
$P_{\alpha'',\alpha'''}(i\omega_n;\vec{\tilde{r}},\vec{\tilde{r}}\,';
\vec{r}\,')$ from the left, apply the operator 
$i\omega_n\delta_{\alpha'',\alpha}\delta_{\alpha''',\alpha'}+[H_0]_{
\alpha'',\alpha''';\alpha,\alpha'}$ to 
$\tilde{P}_{\alpha'',\alpha'''}(i\omega_n;$ $\vec{\tilde{r}},\vec{\tilde{r}}\,';
\vec{r}\,')$ from the right, and obtain  
\bqa
&\Bigl\{i\omega_m\delta_{\alpha,\alpha''}\delta_{\alpha',\alpha'''}-
\Bigl(H_0\Bigl[\vec{\tilde{r}},\frac{1}{i}\frac{\partial}{\partial
\vec{\tilde{r}}};\vec{\tilde{r}}\,',\frac{1}{i}\frac{\partial}{\partial
\vec{\tilde{r}'}}\Bigr]- e^2\frac{1}{|\vec{\tilde{r}}-\vec{\tilde{r}}\,'|}
\Bigr)_{\alpha,\alpha';\alpha'',\alpha'''}\Bigr\} \cr
& \qquad P_{\alpha'',\alpha'''}(i\omega_m;\vec{\tilde{r}},\vec{\tilde{r}}\,';
\vec{r}\,')=\langle c,\alpha|\vec{p}|v,\alpha'\rangle
\delta(\vec{\tilde{r}}-\vec{r}\,')\delta(\vec{\tilde{r}}\,'-\vec{r}\,')
\label{B-15} \cr 
& \sum_{\alpha''}\sum_{\alpha'''}\tilde{P}_{\alpha'',\alpha'''}
(i\omega_m;\vec{\tilde{r}},\vec{\tilde{r}}\,';\vec{r}\,')
\Bigl\{i\omega_m\delta_{\alpha'',\alpha}\delta_{\alpha''',\alpha'}+
\Bigl(H_0\Bigl[\vec{\tilde{r}},\frac{1}{i}\frac{VC{\partial}}{\partial
\vec{\tilde{r}}};\vec{\tilde{r}}\,',\frac{1}{i}\frac{\vec{\partial}}{\partial
\vec{\tilde{r}'}}\Bigr] \cr 
& \qquad -\frac{ e^2}{|\vec{\tilde{r}}-
\vec{\tilde{r}}\,'|}\Bigr)_{\alpha'',\alpha''';\alpha,\alpha'}\Bigr\}=
\langle v,\alpha'|\vec{p}|c,\alpha\rangle
\delta(\vec{\tilde{r}}-\vec{r}\,')\delta(\vec{\tilde{r}}\,'-\vec{r}\,').
\label{B-16}
\enqa
We may then solve the auxiliary functions from Eqs. (\ref{B-15}) and (\ref{B-16})
by the Green's function method to get 
\bq
P_{\alpha,\alpha'}(i\omega_m;\vec{r},\vec{r};\vec{r}\,')=
\sum_{\beta,\beta'}G_{\alpha,\alpha';\beta,\beta'}
(i\omega_m;\vec{r},\vec{r};\vec{r}\,'\vec{r}\,')
\langle c,\beta|\vec{p}|v,\beta'\rangle
\label{B-17}
\enq
with
\bq
G_{\alpha'',\alpha''';\beta,\beta'}(i\omega_m;\vec{\tilde{r}},
\vec{\tilde{r}'};\vec{r},\vec{r}\,')=
\sum_{\lambda,s}\sum_{\lambda',s'}\sum_n
\frac{\psi^{(\lambda,s;\lambda',s';n)}_{\alpha'',\alpha'''}
(\vec{\tilde{r}},\vec{\tilde{r}}\,')
\psi^{\ast(\lambda,s;\lambda',s';n)}_{\beta,\beta'}(\vec{r},\vec{r}\,')}
{i\omega_m-E^{(\lambda,s;\lambda',s')}_n},
\label{B-18}
\enq
as well as
\bq
\tilde{P}_{\alpha,\alpha'}(i\omega_m;\vec{r},\vec{r};\vec{r}\,')=
\sum_{\beta,\beta'}\langle v,\beta'|\vec{p}|c,\beta\rangle
\tilde{G}_{\beta,\beta';\alpha,\alpha'}
(i\omega_m;\vec{r}',\vec{r}';\vec{r},\vec{r})
\label{B-19}
\enq
with
\bq
\tilde{G}_{\beta,\beta';\alpha'',\alpha'''}(i\omega_m;\vec{r},\vec{r}\,';
\vec{\tilde{r}},\vec{\tilde{r}}\,')=\sum_{\lambda,s}\sum_{\lambda',s'}\sum_n
\frac{\psi^{(\lambda,s;\lambda',s';n)}_{\beta,\beta'}
(\vec{r},\vec{r}\,')\psi^{\ast(\lambda,s;\lambda',s';n)}_{\alpha'',\alpha'''}
(\vec{\tilde{r}},\vec{\tilde{r}}\,')}
{i\omega_m+E^{(\lambda,s;\lambda',s')}_n}
\label{B-20}
\enq
in which
$\psi^{(\lambda,s;\lambda',s';n)}_{\alpha,\alpha'}(\vec{r},\vec{r}\,')$
and $E^{(\lambda,s;\lambda',s')}_n$ are exactly those introduced in
Eq.(\ref{m26}). By substituting Eqs. (\ref{B-17})-(\ref{B-20}) into Eq.(\ref{B-14}), we obtain Eq.(\ref{m24}) as desired.
\end{document}